\RequirePackage{fix-cm}
\documentclass[smallextended]{svjour3}
\usepackage{graphicx}
\usepackage{natbib}
\usepackage{url}
\journalname{Computational Statistics}

\newcommand{\libeemd}{\texttt{libeemd}}
\newcommand{\pyeemd}{\texttt{pyeemd}}
\newcommand{\Rlibeemd}{\texttt{Rlibeemd}}
\newcommand{\sourceurl}{\url{https://bitbucket.org/luukko/libeemd}}

\begin{document}

\title{Introducing \libeemd: A program package for performing the ensemble
empirical mode decomposition}
\titlerunning{Introducing \libeemd: A program package for EEMD}

\author{P. J. J. Luukko \and J. Helske \and E. R\"as\"anen}

\institute{%
P. J. J. Luukko \at
Nanoscience Center, University of Jyv\"askyl\"a, FI-40014, Finland\\
\email{perttu.luukko@iki.fi}
\and
J. Helske \at
Department of Mathematics and Statistics, University of Jyv\"askyl\"a, FI-40014, Finland
\and
E. R\"as\"anen \at
Department of Physics, Tampere University of Technology, FI-33101, Finland
}

\date{Received: date / Accepted: date}

\maketitle

\begin{abstract}
The ensemble empirical mode decomposition (EEMD) and its complete variant (CEEMDAN) are adaptive, noise-assisted
data analysis methods that improve on the ordinary empirical mode decomposition
(EMD). All these methods decompose possibly nonlinear and/or nonstationary time
series data into a finite amount of components separated by instantaneous
frequencies. This decomposition provides a powerful method to look into the
different processes behind a given time series data, and provides a way to
separate short time-scale events from a general trend.

We present a free software implementation of EMD, EEMD and CEEMDAN and give an overview
of the EMD methodology and the algorithms used in the decomposition. We release
our implementation, \libeemd, with the aim of providing a user-friendly, fast,
stable, well-documented and easily extensible EEMD library for anyone interested
in using (E)EMD in the analysis of time series data. While written in C for
numerical efficiency, our implementation includes interfaces to the Python and R
languages, and interfaces to other languages are straightforward.

\end{abstract}

\section{Introduction}

Empirical mode decomposition (EMD) is a method for decomposing and analyzing
time series data which may be nonlinear and/or nonstationary \citep{emd}. The
EMD procedure decomposes the input signal (the time series data) into a
collection of intrinsic mode functions (IMFs), which are simple oscillatory
modes with meaningful instantaneous frequencies, and a residual trend. When
combined with Hilbert spectrum analysis (HSA) to extract the instantaneous
frequencies of the IMFs, EMD becomes a powerful tool for analyzing and
characterizing the underlying processes behind any given time series data. The
combination of EMD with HSA is often called the Hilbert-Huang transform~(HHT)
\citep{emdbook}.

While originally a data analysis tool for Earth sciences (see, e.g.,
\cite{emdreview} and references therein), the generality of EMD and HHT has
resulted in applications in fields ranging from medicine~\citep{emdmedicine} to
finance \citep{emdfinance}. It can be used for speech
recognition~\citep{speech} and detrending energy-level spectra of quantum
systems~\citep{emdunfold}. Given its versatility, it is likely that EMD will
continue to find new problems to solve in all fields of science dealing with
empirical data. Hopefully future studies will also give EMD a more
well-understood mathematical foundation, which is still somewhat lacking
\citep{emdreview}.

The original EMD algorithm has also resulted in many derivative algorithms
which improve on the original design and adapt the algorithm to more specific
uses. A recent improvement from the original authors of EMD is ensemble
empirical mode decomposition (EEMD) in which additional noise is used to better
separate different frequency scales into different IMFs without using
subjective selection criteria \citep{eemd}. The original EEMD method is not a
complete decomposition, since the original signal cannot be exactly recovered by
adding together its EEMD components. Instead, a more recent variant CEEMDAN
(Complete EEMD with Adaptive Noise) by~\cite{ceemdan} achieves completeness while
improving the algorithm's robustness against noisy input
signals~\citep{ceemdan2}.

The purpose of this work is to provide a fast, generic, well-documented and
easily accessible implementation of EMD, EEMD and CEEMDAN for anyone interested in using
them in their research. Our code library will also hopefully serve as a basis
for implementing existing and future derivatives of EMD. To foster the use and
further development of our program we release it under a free software license.
The source code of our program can be freely downloaded from \sourceurl.

\section{Algorithms}

\subsection{Empirical mode decomposition}

The basics of the empirical mode decomposition algorithm are documented well in
the literature \citep{emd,emdreview,emdperf}, but for completeness we
recapitulate the general idea. The target of the EMD procedure is to decompose
a signal -- in the spirit of the Fourier series -- into a sum of simple
components. In contrast to the Fourier series, these components are not
required to be simple sinusoidal functions, but they are required to have
meaningful \emph{local} frequencies. The components are called intrinsic mode
functions (IMFs), and the requirement of a meaningful local frequency is
enforced by requiring two conditions \citep{emd}: (1)~the number of zero
crossings and the number of local extrema of the function must differ by at
most one and (2) the ``local mean'' of the function is zero. What is meant by a
local mean in this case is elaborated below.

\begin{figure}
	\centering
	\includegraphics{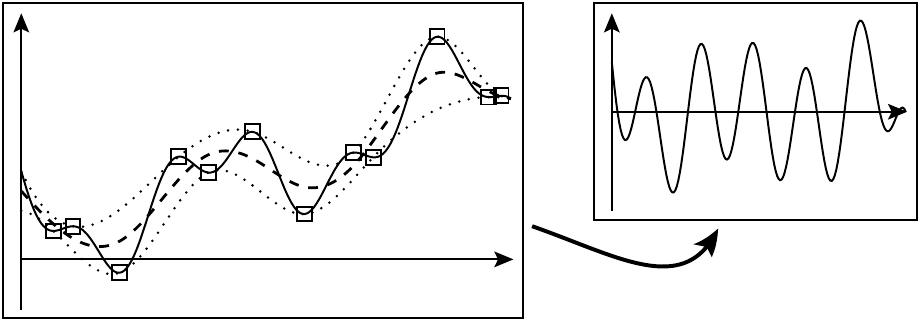}
	\caption{The sifting procedure isolates oscillations around the ``local
	mean'' of the signal. First, on the left panel, local extrema (open
	squares) of the signal (solid line) are located. Then, upper and lower
	envelopes (dotted lines) of the signal are formed by connecting the local
	maxima and minima, respectively, with a spline function. The mean of these
	envelopes (dashed line) is designated as the local mean of the signal. By
	subtracting the local mean from the original signal, oscillations with high
	local frequency are isolated into a new, simpler and more symmetric signal,
	shown on the right panel.}
	\label{fig:siftingdemonstration}
\end{figure}

At the heart of EMD is the sifting procedure, which extracts the simple
oscillatory components (the IMFs) from the original signal. First, upper and
lower envelopes of the signal are computed by finding the local extrema of the
signal. To construct the upper (lower) envelope, the local maxima (minima) are
connected by a smooth interpolation -- typically a cubic spline. Then the mean
of these envelopes is designated as a ``local mean'' of the signal, which can
be used as a reference that separates the lower frequency oscillations in the
signal (the part included in the local mean) from the highest frequency
oscillations (the oscillations around the local mean). This procedure is shown
schematically in Fig.~\ref{fig:siftingdemonstration}. Note that the
separation into ``high frequency'' and ``low frequency'' is now based purely on
the rapidity of oscillations in the original signal. Also, what is considered
high frequency in one part of the signal can be low frequency in another part,
since the local mean can oscillate wildly in some part of the data and change
slowly in other parts.

By subtracting the local mean from the original signal, we can separate the
high (local) frequency oscillations from the rest of the signal. However, this
subtraction can create new local extrema, foiling the requirements set for an
IMF. To actually recover the highest frequency IMF component, the sifting
procedure is applied again and again, until some stopping criterion is
fulfilled and we are left with a sufficiently pure IMF. The choices of stopping
criteria are discussed further in Sec.~\ref{sec:stopping}. After the first IMF
is obtained, it can be subtracted from the original signal, and the procedure
outlined above can be used to extract the IMF with the second-highest local
frequency. This can be repeated until the residual signal is monotonous, and no
further IMFs can be extracted. This residual, possibly together with the lowest
frequency IMFs, can be used to represent the intrinsic trend of the data.
Detecting and removing such a trend is alone a useful application of EMD for
several purposes \citep{emddetrend,emdunfold}.

EMD separates different frequency scales of the signal into separate IMFs, but
it is not guaranteed that -- when analyzing a data from some natural process --
each IMF represents a physical time scale of the process. Often ranges of IMFs
need to be added together to extract information pertaining to a single
natural time scale \citep{eemd}, and some IMF components may represent the
properties of measurement noise instead of the underlying physical process. To
assist in selecting the IMFs with a physical meaning \cite{emdsign} have
constructed a statistical significance test which compares the IMFs against a
null hypothesis of white noise.

\subsection{Detection of extrema and zero slopes}
\label{sec:extrema}

\begin{figure}
	\centering
	\includegraphics{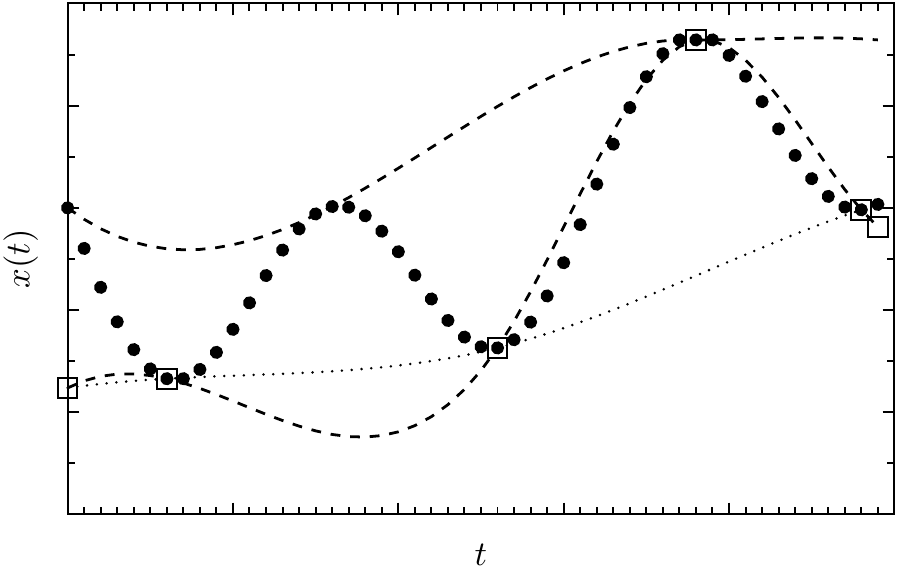}
	\caption{Plot highlighting the problem caused by equal consecutive data
	  points in the reference EEMD implementation by \cite{eemd}. The filled
	  circles show an artificial signal $x(t)$, in which three data points near
	  the top of an oscillation happen to be equal within sample precision.
	  The squares and the dashed lines show the local minima and the envelope
	  splines, respectively, as found by the reference EEMD implementation.
	  Because the middle one of the equal data points is considered both a
	  maximum and a minimum, the lower envelope spline shoots up sharply, no
	  longer representing a good lower envelope for the signal. The lower
	envelope produced by \libeemd\ is shown as a dotted line}
	\label{fig:extremaproblem}
\end{figure}

The local extrema of the data sequence, which are needed to form the upper and
lower envelopes, can of course be found by simply comparing consecutive data
points. However, some care is needed in the handling of zero slopes, i.e.,
stretches of strictly equal data points in the sequence. The reference EEMD
library by \cite{eemd} considers consecutive equal data points to be
\emph{both} maxima and minima, which causes the upper and lower envelopes to
meet at these points. This is probably an unwanted feature since, e.g., a small
deviation at the highest point of a large oscillation can cause the topmost
data points to be equal, causing in turn the \emph{lower} envelope to jump
sharply to meet the upper envelope at the local maximum, as illustrated
in~Fig.~\ref{fig:extremaproblem}.

Naturally, exactly equal data points in the original signal are likely to be a
result of artificial data or low sample accuracy, but such zero slopes can also
occur in the intermediate steps of the EMD procedure, especially when the
number of samples is low.

To ameliorate these issues with zero slopes, we choose a different convention
for the extrema detection. For a point to be considered a local maximum, the
slope before the data point has to be strictly positive, and the slope after
the data point strictly negative. However, if there is a region of zero slope
between the positive and the negative slopes, the \emph{center} of the zero
slope region is designated as a single maximum. Likewise, valleys with a flat
bottom are considered to have a single minimum at the center of the flat
region. This convention is also used by another implementation
by~\cite{ceemdan}.

\subsection{Spline envelopes}

There are several ways to interpolate the local extrema to form the upper and
lower envelopes. A cubic spline is considered by most to be the best trade-off
between smooth envelopes and simplicity \citep{emdreview}, but there are
several different kinds of cubic splines. In their review \cite{emdperf}
consider the so-called natural cubic spline the best choice for data from water
waves. We adopted the convention used by the reference EEMD library by
\cite{eemd} and used cubic splines with the ``not-a-knot'' end conditions.
These splines are somewhat more computationally intensive to calculate, but
unlike with the ``natural'' end conditions, the splines are not required to
have zero curvature at the ends of the data, which we consider to be an
artificial limitation. For computing the splines we use the algorithm described
by \cite{numalg}. As the reference EEMD implementation, our code falls back to
linear interpolation (for $N=2$) or polynomial interpolation (for $N=3$) if the
number of extrema~$N$ is too small for a cubic spline.

More recently effort has been made into using B-splines for the envelopes to
establish a firmer mathematical basis for EMD \citep{bsplines}. In addition,
piecewise cubic Hermite interpolation polynomials (PCHIPs) have been suggested
to replace the usual cubic splines to better interpolate highly nonstationary
signals \citep{pchip}. Both of these are considered to be useful alternatives
to be added in the future to \libeemd.

\subsection{End effects}
\label{sec:endeffects}

While cubic splines are very good at interpolating data at the interior of the
data sequence, they perform worse at the ends of the data. This is of course a
problem shared by interpolation in general -- at the ends of the data the
interpolation algorithm can only work with the ``neighborhood'' of data points
on one side of the end, since the other side is missing. This turns the
interpolation problem into an extrapolation one -- how to predict how the data
would behave before the start of the sequence? If left unattended, the errors
of the spline interpolation can cause the ends of the computed IMFs to become
corrupted, and as the sifting procedure continues these errors propagate to the
interior to the data.

Several ways have been proposed to mitigate the end effects by adding
artificial extrema to the ends of the data, such as simple wave forms defined
by the extrema near the end \citep{emd}. We have adopted the method described
by \cite{eemd}, where additional extrema are added to the ends of the data by
linear extrapolation of the previous two extrema. However, if the extrapolated
extremum is less extremal than the last data point, the value of the last data
point is used as an additional extremum instead. This method successfully
reduces the end effects while avoiding the possible complications of more
complex data extrapolation.

\subsection{Stopping criteria}
\label{sec:stopping}

Many ways to decide how many times a signal is sifted to produce a single IMF
have been proposed. The original algorithm by \cite{emd} used a Cauchy-like
criterion, in which the sifting is stopped when the normalized square
difference of results from consecutive iterations is sufficiently small.
However, this method is vulnerable to sifting the signal too much so that it
becomes a frequency-modulated signal only and all amplitude variation is lost.
It is also not guaranteed that the final result will fulfill the requirements
of an IMF. Therefore \cite{Snumber} proposed a simpler stopping criteria, in
which iteration is stopped when the number of zero crossings and extrema differ
at most by one \emph{and} that these numbers stay the same for~$S$ consecutive
iterations. This criterion was extensively studied by \cite{emdconv} and the
optimal range for the $S$-number was found to be from~3 to~8. Our code
supports\footnote{In rare cases the finite precision of computer arithmetic can
cause the number of zero crossings or extrema to get stuck oscillating between two
consecutive numbers. To avoid an endless loop or extreme oversifting in this
case our implementation relaxes the latter condition so that a case where only
one of the numbers changes by 1 is still considered stable. This change does
not affect the normal operation of EMD.} using this stopping criterion for the
siftings with a default value of~$S=4$. In addition, a maximum number of total
iterations can be set to prevent oversifting. By only setting the maximum
number of iterations, EMD can be performed with a fixed number of iterations,
which is sometimes preferred as described by \cite{eemd}.

\subsection{Ensemble EMD and CEEMDAN}

A recently demonstrated improved variant of the EMD method is Ensemble EMD
(EEMD), in which EMD is performed on an ensemble of initial signals, each
perturbed by low-amplitude white noise \citep{eemd}. The noise helps the sifting
process to avoid mode mixing and to provide more robust and physically
meaningful IMFs. In the end the average of the results is designated as the
true final result, and thus the direct effect of the noise is canceled out.
Computing the EMD of a large ensemble of signals is computationally more
intensive, but this difference in computation time can be reduced significantly
since the separate ensemble members can be computed in parallel.

Because the added noise does not completely cancel out in the averaging process
for any finite ensemble size, EEMD is no longer a strictly complete
decomposition. This issue has been fixed in a EEMD variant called CEEMDAN
\citep{ceemdan}. In CEEMDAN, the averaging over all ensemble members is carried
separately for each IMF component. By changing the order of averaging over the
ensemble and extracting the next IMF, at each point of the decomposition
procedure the current residual together with the already extracted IMFs sums
exactly (or up to numerical precision) to the original signal. This small
change also seems to improve the algorithm's efficiency in recovering an
underlying tone from an already noisy input signal \citep{ceemdan2}.

\section{Implementation details}

\subsection{Low-level C library}

The low-level computational methods of \libeemd\ are written in
standards-compliant C99 for numerical efficiency and portability. The C
interface to \libeemd\ is documented in the header file \texttt{eemd.h}
distributed with \libeemd. All the methods for computing EMD, EEMD or CEEMDAN can be
used directly from C, but we focus on providing interfaces so that \libeemd\
can be accessed from more high-level languages which are better suited for data
analysis. We provide complete interfaces to the Python and R languages, which are
excellent modern languages for data analysis tools and other scientific
software. In the future, interfaces to other languages will be
considered.

For random number generation and basic linear algebra we use routines provided
by the GNU Scientific Library (GSL) \citep{gsl}. For faster EEMD and CEEMDAN computations
with large ensembles the separate ensemble members are handled in parallel with
OpenMP. More elaborate parallelization schemes, which benefit large,
multidimensional EEMD calculations, also exist in the
literature~\citep{parallelmeemd}. The library can also be compiled in
single-thread mode if parallelization is not desired.

\subsection{Python interface}

The \libeemd\ distribution includes a Python interface called \pyeemd. This
interface allows the routines in \libeemd\ to be called directly from Python,
using standard Numpy \citep{numpy} arrays for input and output data. Via
\pyeemd\ the routines provided by \libeemd\ can be accessed with a minimal
amount of overhead code. To give a short example, if we have an input signal
stored in a 1D Numpy array (or some other Python sequence type) \texttt{input},
we can decompose it with EEMD using the code:
\begin{verbatim}
import pyeemd
imfs = pyeemd.eemd(input)
\end{verbatim}
Now \texttt{imfs} is a $M\times N$ array, where $N$ is the length of
\texttt{input}, and the IMFs it decomposed to are stored in the~$M$ rows of
\texttt{imfs}, the last row being the residual. The individual IMFs can then be
manipulated easily with the numerous arithmetic tools provided by Numpy. For
completing the Hilbert-Huang transform, the Hilbert transformation routine
provided by the Scipy package (\texttt{scipy.fftpack.hilbert}) can be used. The
IMFs can be visualized by any of the several plotting libraries available to
Python, but we also provide a simple helper routine
(\texttt{pyeemd.utils.plot\textunderscore imfs}) for quick visualization of the
results.

The stopping criterion for the decomposition can be set by using the optional
parameters \texttt{S\textunderscore number} and \texttt{num\textunderscore
siftings}, corresponding to the $S$-number criterion and a maximum number of
siftings, respectively, as described in Sec.~\ref{sec:stopping}. The default values are
\texttt{S\textunderscore number=4} and \texttt{num\textunderscore siftings=50}.
Other optional
parameters that can be used to influence the EEMD decomposition are
\texttt{ensemble\textunderscore size} and \texttt{noise\textunderscore
strength}. The parameter \texttt{ensemble\textunderscore size} is the size of
the EEMD ensemble and it defaults to~250. The magnitude of the Gaussian white
noise added to the input in EEMD is controlled by \texttt{noise\textunderscore
strength}, which represents the standard deviation of the Gaussian random
numbers used as the noise, \emph{relative} to the standard deviation of the
input. It defaults to setting the noise to have~$0.2$ times the standard
deviation of the signal. This value was suggested by \cite{eemd} and shown to
be a good default value for many cases by \cite{ceemdan2}. Computing a simple
EMD decomposition can be done by
setting \texttt{ensemble\textunderscore size} to~1 and
\texttt{noise\textunderscore strength} to~0, but we also provide as a shortcut
the routine \texttt{pyeemd.emd}.

The \pyeemd\ distribution also includes unit tests which can be used to ensure
that the code is working as intended. These tests can also be used to ensure
that our code reproduces the results obtained by the reference EEMD
implementation, except for the case of equal data points described in
Sec.~\ref{sec:extrema}. For more documentation of the use and internals of
\pyeemd\ please see the documentation files and source code distributed with
\pyeemd, and the online documentation at \url{http://pyeemd.readthedocs.org/}.

\subsection{R interface}

In addition to a Python interface, we have written an interface to R in a
form of a complete R package \Rlibeemd. A stable version of the package is
available at CRAN\footnote{\url{http://cran.r-project.org/web/packages/Rlibeemd/index.html}}, and the latest
development version is also available at
GitHub\footnote{\url{https://github.com/helske/Rlibeemd}}. With help of
\texttt{Rcpp} package \citep{RcppA, RcppB} the \libeemd\ C library is integrated
into R with minimal overhead. Input data for decomposition can be a numeric
vector or an object which can be coerced to such. Output data is converted to a
time series object of class \texttt{mts} for easier plotting and further
analysis. We present a short example using the UK gas consumption dataset which
is a part of the base R dataset collection. We perform a CEEMDAN decomposition
for logarithmic consumption using default values of the \texttt{ceemdan}
function, and plot the resulting IMFs:

\begin{verbatim}
library(Rlibeemd)
# logarithmic demand of quarterly UK gas consumption
imfs <- ceemdan(log(UKgas))
# set the layout and marginals for the plots
par(mfrow = c(2, 2), mar = c(2, 4, 1, 1), oma = c(1, 1, 3, 1), 
    xaxs = "i")
plot(log(UKgas), ylab = "Observations")
plot(imfs[, 1], ylab = "Seasonal")
# Sum IMFs 2 to 5 as one irregular component
plot(ts(rowSums(imfs[, 2:5]), start = start(UKgas), freq = 4), 
     ylab = "Irregular")
plot(imfs[, 6], ylab = "Trend")
\end{verbatim}

\begin{figure}
	\centering
	\includegraphics[width=\textwidth]{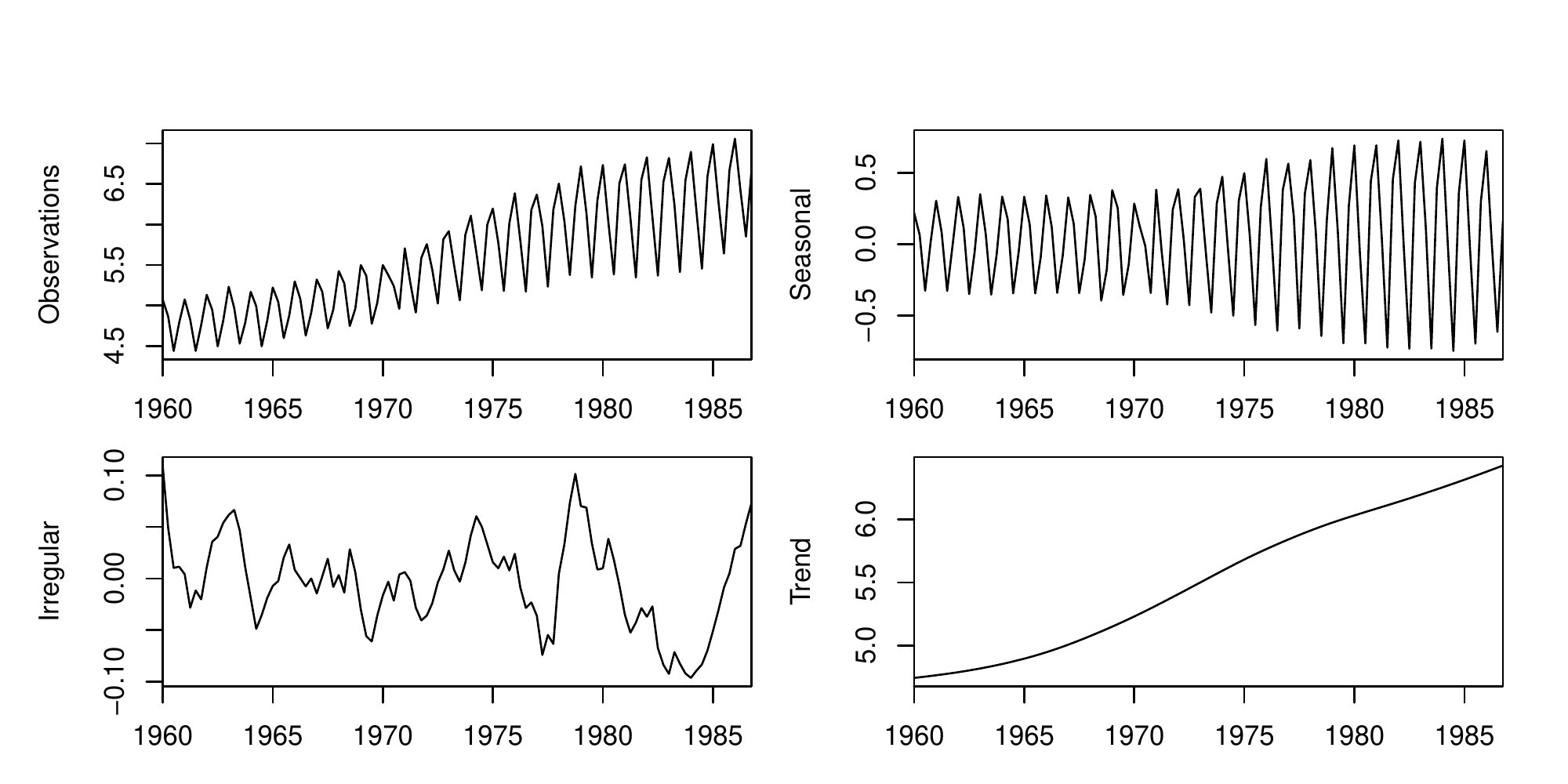}
	\caption{Example plot produced using \Rlibeemd, showing the various
	components extracted from UK gas consumption with CEEMDAN.}
	\label{fig:ukgas}
\end{figure}

CEEMDAN extracts five IMFs and the residual, which captures the underlying
trend. The first IMF contains the seasonal effect whereas rest of the IMFs
contain the irregular part. In the example code, the IMFs number 2 to 5 are combined by taking the sum over the IMFs at each time point. The resulting irregular component is then defined as a time series object for proper plotting behavior. The resulting plot can be seen in
Fig.~\ref{fig:ukgas}. More examples can be found from the documentation
distributed with the \Rlibeemd\ package.

\section{Comparison to other implementations}

Several established EMD, EEMD and CEEMDAN implementations are publicly
available. Of these the most commonly used ones are the reference EEMD
implementation by \cite{eemd}, the EMD toolbox by P. Flandrin's
group\footnote{\url{http://perso.ens-lyon.fr/patrick.flandrin/emd.html}}, the
original CEEMDAN implementation by \cite{ceemdan} (also available from P.
Flandrin's home page), the \texttt{EMD} R package by \cite{Remd}, and the \texttt{hht} R package
by \cite{Rhht} containing EEMD and CEEMDAN implementations. There is also a rudimentary Python implementation of EMD/EEMD
included in the Python Time Series Analysis package
(PTSA)\footnote{\url{http://ptsa.sourceforge.net}} by the Ohio State University
Computational Memory Lab. We are also aware of an EEMD code for Matlab by
\cite{feemd}, but it is only available in binary form. While the algorithms
implemented in \libeemd\ are implemented also by others (with possible minor
variations such as discussed in Sec.~\ref{sec:extrema}), our implementation
brings several benefits, which we outline in the following.

Except the packages written for R and the PTSA, all other implementations
mentioned here are written for Matlab. It is very
difficult to use routines written in Matlab from other languages, and using
Matlab requires the purchase of a license. Likewise, R codes are mostly only
usable on R, and Python codes in projects using Python (some commercial statistical software such as SAS and SPSS
support R or Python plug-ins). Having a stable C
library is very useful for expanding the user base of EMD, since almost every
high-level language uses C as the low-level language for writing extension
modules. The C interface of \libeemd\ has been designed to make the work of
interface builders as easy as possible by only using standard C data types in
the public interface. Having the algorithms written in a low-level language
such as C also brings immediate and substantial speed improvements over
implementations using only interpreted languages.

We would also like to highlight the importance of clearly written,
well-documented and thoroughly tested code, as well as modern programming
practices and proper version control. We have taken effort in making the code
of \libeemd\ easily readable and modifiable by others, making it easier for
other people to use \libeemd\ as a basis for not only new applications of EMD,
but also new variants of the underlying algorithms. The source code of
\libeemd\ is hosted on Bitbucket\footnote{\sourceurl}, making it easy for users
to track changes in the program, report issues, and discuss and publish
improvements to the program. Being published under an open-source license,
\libeemd\ protects the users' right to modify the code, fostering the
development of better software.

Besides the original CEEMDAN implementation and the R package \texttt{hht}, to our knowledge there are
currently no other publicly available implementations of CEEMDAN. Our
implementation thus also provides the first C and Python implementations of
CEEMDAN, as well as the first parallelized CEEMDAN implementation.

There are also implementations that focus on computing EMD-like decompositions
in real time from measurement data, often focusing on specific kinds of data
and using specialized hardware and/or GPU acceleration. Our implementation
focuses on providing a free and generic software implementation of the EMD
algorithms for offline data analysis, so we have not included these
implementations in our comparison.

\subsection{Performance comparison}

As mentioned before, writing the core algorithm in a low-level language brings
a substantial improvement in numerical performance. To provide an example in
the case of R, we compare briefly the performance of the R package
\texttt{hht}~\citep{Rhht} and our \Rlibeemd. As input data, we use ECG
(electrocardiogram) data from the MIT-BIH Normal Sinus Rhythm
Database\footnote{\url{http://www.physionet.org/cgi-bin/atm/ATM}}, which was
also used in the original article of \cite{ceemdan}. We use a value of 0.2 as
the relative standard error of added noise, and the S-number stopping criterion
with $S = 4$ together with a maximum number of siftings of 50. Length of the
signal is varied from 100 to 2000, and ensemble size is varied from 50 to
500. The maximum number of IMFs to extract (including the residual) is set to
six.

\begin{figure}
	\centering
	\includegraphics{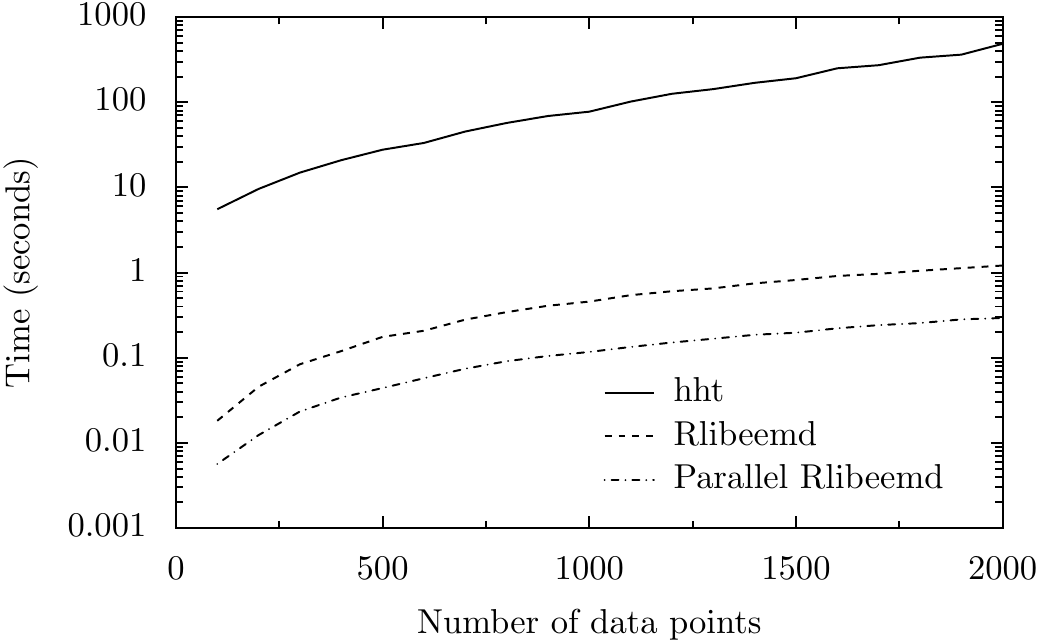}
	\caption{Performance comparison of the R packages \texttt{hht} (function
	\texttt{CEEMD}) and \Rlibeemd\ (function \texttt{ceemdan}) for varying
	number of input data points. The y-axis shows the mean computation time
	required to run the CEEMDAN decomposition function. The implementation from
	\Rlibeemd\ is approximately two orders of magnitude faster than
	\texttt{hht}, and it can be sped up further with parallelization.}
	\label{fig:varyingN}
\end{figure}

Benchmarking was performed on a system with an Intel Quad-Core i7-4770
3.40\,GHz CPU and~16 GB of RAM running on~64 bit Microsoft
Windows~7 Enterprise platform. Figure~\ref{fig:varyingN} shows how the required
computation time for CEEMDAN scales with the number of data points. Note that
the y-axis is logarithmic in order to deal with the different scales of
performance. The performance of \texttt{Rlibeemd} is clearly superior to
\texttt{hht}, which is approximately two orders of magnitude slower than
non-parallerized \Rlibeemd. The parallerized version of \Rlibeemd\ is
approximately four times faster than the non-parallerized \Rlibeemd, which is
the expected speedup for a quad-core CPU. Similar results were obtained also
when the ensemble size was varied instead of the number of data points, and
when comparing results for the EEMD algorithm. 

It should be noted that in addition to extracting the IMFs, the \texttt{CEEMD}
function of \texttt{hht} also performs a Hilbert transform on the IMFs in order
to obtain their instantaneous frequencies. However, this additional step
should have only a marginal effect to overall computational time, as the Hilbert
transform only needs to be executed once for the final IMFs.

We also compared the performance of \Rlibeemd\ and the R package
\texttt{EMD}~\citep{Remd} in performing the ordinary EMD decomposition. With
the same overall parameters as the CEEMDAN comparison, the \texttt{EMD} package
was found to be approximately three orders of magnitude slower than \Rlibeemd.

\section{Conclusions}

We have presented a free software code library which implements the ensemble
empirical mode decomposition (EEMD), of which the regular empirical mode
decomposition (EMD) is a special case, and its complete variant CEEMDAN. Since
our library is implemented in~C,
it can be readily interfaced with a variety of high-level languages for
inclusion into existing data analysis software. As an example of this we have
provided complete Python and R interfaces. By implementing the algorithm in C and providing interfaces to high-level languages the implementation retains the unmatched speed of a low-level language while gaining the ease of use and flexibility of higher level languages.
Our implementation corrects a minor issue
with equal data points in the reference EEMD implementation by~\cite{eemd}, includes the first optimized and parallel implementation of CEEMDAN, and
provides a solid and well-documented basis for existing and future improvements
of the (E)EMD method.

\begin{acknowledgements}
This work was supported by the Finnish Cultural Foundation, the Emil Aaltonen Foundation, the Academy of
Finland, and the European Community's FP7 through the CRONOS project, Grant
Agreement No.~280879. The authors wish to thank N.~E.~Huang for useful
discussions.
\end{acknowledgements}

\bibliographystyle{spbasic}      
\bibliography{libeemd-paper}

\end{document}